\documentstyle[graphics]{mn}

\newlength{\colwidth}
\newcommand{\ion}[2]{\hbox{#1\,{\sc #2}}}
\newcommand{\HI}{\ion{H}{i}}
\newcommand{\HII}{\ion{H}{ii}}
\newcommand{\HeI}{\ion{He}{i}}
\newcommand{\HeII}{\ion{He}{ii}}
\newcommand{\HeIII}{\ion{He}{iii}}
\newcommand{\vpfit}{{\sc {VPFIT}}} % Need the extra brackets.
\newcommand{\hydra}{{\sc {HYDRA}}}
\newcommand{\bpar}{$b$-parameter}
\newcommand{\bpars}{$b$-parameters}

\newcommand{\cm}{\rm cm}
\newcommand{\kms}{{\rm km}\,{\rm s}^{-1}}
\newcommand{\K}{\rm K}

\newcommand{\lya}{Ly$\alpha$}
\newcommand{\lyb}{Ly$\beta$}

\title[Measuring the equation of state of the IGM]%
      {Measuring the equation of state of the intergalactic medium}
\author[J.~Schaye et al.] %
	{Joop Schaye,$^1$ Tom Theuns,$^{2,1}$ Anthony Leonard$^3$ and 
	George Efstathiou$^1$\\
	$^{1}$Institute of Astronomy, Madingley Road, Cambridge CB3 0HA\\
	$^{2}$Max-Planck-Institut f\"ur Astrophysik, Postfach 1523,
	85740 Garching, Germany\\
	$^{3}$Department of Physics, Astrophysics, University of
	Oxford, Keble Road, Oxford OX1 3RH}

\begin{document}

\maketitle

\begin{abstract}
Numerical simulations indicate that the smooth, photoionized
intergalactic medium (IGM) responsible for the low column density
\lya\ forest follows a well defined temperature-density relation,
which is well described by a power-law
$T=T_0(\rho/\bar{\rho})^{\gamma-1}$. We demonstrate that such an
equation of state results in a power-law cutoff in the distribution of
line widths ($b$-parameters) as a function of column density ($N$) for
the low column density ($N \la 10^{14.5}\,\cm^{-2}$) absorption
lines. This explains the existence of the lower envelope which is
clearly seen in scatter plots of the $b(N)$-distribution in observed
QSO spectra. Even a strict power-law equation of state will not result
in an absolute cutoff because of line blending and contamination by
unidentified metal lines. We develop an algorithm to determine the
cutoff, which is insensitive to these narrow lines. We show that the
parameters of the cutoff in the $b(N)$-distribution are strongly
correlated with the parameters of the underlying equation of state. We
use simulations to determine these relations, which can then be
applied to the observed cutoff in the $b(N)$-distribution to measure
the equation of state of the IGM. We show that systematics which
change the $b(N)$-distribution, such as cosmology (for a fixed
equation of state), peculiar velocities, the intensity of the ionizing
background radiation and variations in the signal to noise ratio do
not affect the measured cutoff. We argue that physical processes that
have not been incorporated in the simulations, e.g.\ feedback from
star formation, are unlikely to affect the results. Using Monte Carlo
simulations of Keck spectra at $z=3$, we show that determining the
slope of the equation of state will be difficult, but that the
amplitude can be determined to within ten per cent, even from a single
QSO spectrum. Measuring the evolution of the equation of state with
redshift will allow us to put tight constraints on the reionization
history of the universe.
\end{abstract}

\begin{keywords}
cosmology: theory -- intergalactic medium -- equation of state --
hydrodynamics -- quasars: absorption lines
\end{keywords}

\section{INTRODUCTION}
The intergalactic medium (IGM) at high redshift ($z\sim 2$--5) manifests itself
observationally by absorbing light from distant quasars. 
Resonant \lya\ absorption by neutral hydrogen along the line of sight
to a quasar results in a fluctuating \lya\ transmission (optical depth).
Regions of enhanced density give rise to increased absorption and
appear as a forest of absorption lines bluewards of the quasar's \lya\
emission line. The fact that not all the light is absorbed implies
that the IGM is highly ionized \cite{gunn65:gp_effect}, but the time and
origin of the reionization of the gas are still unknown. Since quasars
and young stars are sources of ionizing radiation, the ionization
history of the gas depends on the evolution of the quasar
population and the star formation history of the universe.

Hydrodynamical simulations of structure formation in a
universe dominated by cold dark matter and including an ionizing background,
have been very successful in explaining the properties of the \lya\
forest \cite{cen94:lya_forest,%
zhang95:lya_forest,hernquist96:lya_forest,miralda-escude96:lya_forest,%
theuns98:lowz_lya,theuns98:apmsph,dave99:lowz_lya}. They
show that the low column density ($N \la
10^{14.5}\,\cm^{-2}$) absorption lines arise in a smoothly
varying low-density ($\delta \la 10$) IGM. Since the overdensity is
only mildly non-linear, the physical processes governing this medium
are well understood and relatively easy to model. On large scales the
dynamics are determined by gravity, while on small scales gas pressure
is important. The existence of a simple physical framework and the
abundance of superb observational data make the \lya\ forest an
extremely promising cosmological laboratory (see
Rauch~\shortcite{rauch98:review} for a review). In this paper, we shall
investigate the effect of the thermal state of the IGM on the forest.

For the low-density gas responsible
for the \lya\ forest, shock heating is not important and the gas follows a
well defined temperature-density relation. The competition between
photoionization heating and adiabatic cooling results in a power-law
`equation of state' $T=T_0(\rho/\bar{\rho})^{\gamma-1}$
\cite{hui97:tempdens}. This equation of state depends on cosmology and
reionization history. For models with abrupt reionization, the IGM
becomes nearly 
isothermal ($\gamma \approx 1$) at the redshift of reionization. After
reionization, the temperature at the mean density ($T_0$) decreases while
the slope ($\gamma-1$) increases because 
higher density regions undergo less expansion and increased
photoheating. Eventually, when
photoheating balances adiabatic cooling as the universe expands, the
imprints of the reionization history are washed out and the equation
of state approaches an asymptotic state, $\gamma = 1.62$, $T_0 \propto
\left [ \Omega_bh^2 / \sqrt{\Omega_m h^2} \right ]^{1/1.7}$
\cite{hui97:tempdens}. Since the reionization history of the universe
is still unknown, the physically reasonable ranges for the parameters
of the equation of state are very large ($10^{3.0}\,\K < T_0 <
10^{4.5}\,K$ and  $1.2 < \gamma < 1.7$
\cite{hui97:zeldovich_coldensdistr,hui97:tempdens}).  

The smoothly varying IGM gives rise to a fluctuating optical depth in
redshift space. Many of the optical depth maxima can be fitted quite
accurately with Voigt profiles. The distribution of line widths
depends on the initial power spectrum, the peculiar velocity gradients
around the density peaks and on the temperature of the IGM. However,
there is a lower limit to how narrow the absorption lines can
be. Indeed, the optical depth will be smoothed on a scale determined
by three processes \cite{hui97:bdistr}: thermal broadening, baryon
(Jeans) smoothing and possibly instrumental, or in the case of
simulations, numerical resolution. The first two depend on the thermal
state of the gas. While for high-resolution observations (echelle
spectroscopy) the effective smoothing scale is not determined by the
instrumental resolution, numerical resolution has in fact been the
limiting factor in many simulations (see Theuns et
al.~\shortcite{theuns98:apmsph} for a discussion).

The distribution of line widths is generally expressed as the
distribution of the widths of Voigt profile fits to the absorption
lines, the $b$-parameters. While the first numerical simulations showed good
agreement with the observed $b$-parameter distribution, higher
resolution simulations of the standard cold dark matter model produced
a larger fraction of narrow lines than
observed \cite{theuns98:apmsph,bryan98:lya_numeffects}. Theuns et
al.~\shortcite{theuns98:apmsph} suggested that an increase in the
temperature of the IGM might broaden the absorption lines, while Bryan
et al.~\shortcite{bryan98:lya_numeffects} argued that the most natural
way to broaden the lines is to change the density distribution
directly. Note that increasing the temperature will also change the
density distribution of the gas through increased baryon
smoothing (Theuns, Schaye \& Haehnelt 1999).

Theuns et al.~\shortcite{theuns99:cosmology} showed that changing
the cosmology (lowering $\Omega_m$ from 1.0 to 0.3 and doubling
$\Omega_bh^2$ to 0.025) significantly broadens the absorption lines,
although some discrepancy with observations may remain. 
One way to
increase the temperature further would be to change the reionization
history. Uncertainties in the redshift of He reionization, in
particular, can affect the temperature of the IGM.
Haehnelt \& Steinmetz~\shortcite{haehnelt98:bparam} demonstrated
that different reionization histories result in observable differences
in the $b$-distribution. Other mechanisms that have been proposed to boost
the temperature are photoelectric heating of dust grains
~\cite{nath99:dust_heating}, Compton
heating by the hard X-ray background \cite{madau99:xray} and radiative
transfer effects associated with the ionization of \HeII\ by QSOs in
the optically thick limit \cite{abel99:rad_transfer_igm}.

Unfortunately, the $b$-distribution is not very well suited for
investigating the thermal state of the IGM. Although some of the broad
lines correspond to density fluctuations on scales that are affected,
among other things, by thermal smoothing \cite{theuns99:bigb}, many
are caused by heavy line blending and continuum fitting errors. The
cutoff in the $b$-distribution, on the other hand, will depend mainly
on the temperature of the IGM and is therefore potentially a powerful
statistic. In practice its usefulness is limited because many narrow
lines occur in the wings of broader lines. Such narrow lines are often
introduced by numerical Voigt profile fitting algorithms (such as
\vpfit~\cite{carswell87:vpfit}) to improve the quality of the overall
fit to the quasar spectrum. If the physical structure responsible for
the absorption does not consist of discrete clouds, then the widths of
these blended lines will have no relation to the thermal state of the
gas. Furthermore, the number and widths of these narrow blended lines
depend on the Voigt profile fitting algorithm that is used and on the
signal to noise of the quasar spectrum.

The $b$-parameter distribution is usually integrated over a certain column
density range. Since the $b$-distribution might depend on column density
($N$), more information is contained in the full
$b(N)$-distribution. Scatter plots of the $b(N)$-distribution have been
published for many observed QSO spectra
\cite{hu95:lyaobs,lu96:lya_z=4,kirkman97:lya_obs,kim97:lya_evolution}.
These plots show a clear cutoff at low $b$-parameters. However, this
cutoff is not absolute. There are some narrow lines, especially at low column
densities. Lu et al.~\shortcite{lu96:lya_z=4} and Kirkman \&
Tytler~\shortcite{kirkman97:lya_obs} use Monte Carlo simulations to
show that many of these lines are caused by line blending and noise in
the data. Some contamination from unidentified metal lines is also
expected. 

The cutoff in the $b(N)$-distribution increases slightly with column
density. Hu et al.~\shortcite{hu95:lyaobs} conclude from Monte
Carlo simulations that this correlation is primarily an artifact of
the much larger number of lines at lower column density and the
increased scatter in the $b$-determinations. Kirkman \&
Tytler~\shortcite{kirkman97:lya_obs} however, conclude from similar
simulations that the correlation between the lowest $b$-values and column
density is a real physical effect. (Note that this correlation is
different from the one reported by Pettini et
al.~\shortcite{pettini90:high_resol_lya}. They found a general
correlation between 
the \bpar\ and column density of all lines. This correlation was later
shown to be an artifact of the line selection and fitting
procedure~\cite{rauch93:bNcorr}.) A lower envelope which increases
with column density has also been seen in numerical simulations
\cite{zhang97:lya_forest}.

In this paper we shall demonstrate that the cutoff in the
$b(N)$-distribution is determined by the equation of state of the
low-density gas. Furthermore, we shall show that the cutoff can be
determined robustly and is unaffected by systematics like changes in
cosmology (for a fixed equation of state) and can therefore be used to
measure the equation of state of the IGM.

We test our methods using smoothed-particle hydrodynamic (SPH)
simulations of the \lya\ forest as described by Theuns et
al.~\shortcite{theuns98:apmsph,theuns99:cosmology}. The parameters of
the simulations are summarised in section~\ref{sec:simulations}.  The
relation between the $b(N)$-cutoff and the equation of state is
investigated in section~\ref{sec:relation}, which forms the heart
of the paper.  Section~\ref{sec:algorithm} contains a detailed
description of the procedure used to fit the cutoff in simulated Keck
spectra. Systematic effects are discussed in
section~\ref{sec:systematics}. In section~\ref{sec:montecarlo} we test
the procedure using Monte Carlo simulations. Finally, we summarise and
discuss the main results in section~\ref{sec:discussion}.

\section{SIMULATIONS}

\label{sec:simulations}

We have simulated six different cosmological models, characterised by
their total matter density $\Omega_m$, the value of the cosmological
constant $\Omega_\Lambda$, the rms of mass fluctuations in spheres of
radius 8$h^{-1}$ Mpc, $\sigma_8$, the baryon density $\Omega_bh^2$ and the
present day value of the Hubble constant, $H_0 \equiv 100 h~{\rm km}\,{\rm
s}^{-1}\,{\rm Mpc}^{-1}$. The parameters of these models are summarised in
Table~\ref{tbl:models}. In addition to these models, we
simulated a model that has the same parameters as model Ob, but
with the \HeI\ and \HeII\ heating rates artificially doubled. This may
provide a qualitative model of the heating due to radiative transfer
effects during the reionization of helium
\cite{abel99:rad_transfer_igm}. We will call this model Ob-hot. The 
amplitude of the initial power spectrum is 
normalised to the observed abundance of galaxy clusters at $z=0$, using
the fits computed by Eke, Cole \& Frenk~\shortcite{eke96:sigma8}.
We model the evolution of a periodic, cubic region of the universe
of comoving size $2.5h^{-1}$~Mpc.
\begin{table}
\caption{Models simulated}
\begin{tabular}{llllll}
Model & $\Omega_m$ & $\Omega_\Lambda$ & $\Omega_bh^2$ & $h$ &
$\Gamma_{\HI}$ \\
\hline
S    & 1      &   0    &  0.0125  & 0.5  & HM/2 \\
Sb   & 1      &   0    &  0.025   & 0.5  & HM   \\
O    & 0.3    &   0    &  0.0125  & 0.65 & HM/2 \\
Ob   & 0.3    &   0    &  0.025   & 0.65 & HM   \\  
%Ob-hot  & 0.3    &   0    &  0.025   & 0.65 & HM   \\
L    & 0.3    &   0.7  &  0.0125  & 0.65 & HM/2 \\
Lb   & 0.3    &   0.7  &  0.025   & 0.65 & HM   \\
\hline
\end{tabular}
\label{tbl:models}
\end{table}

The code used is adapted from the \hydra\ code of Couchman et
al.~\shortcite{couchman95:hydra}, which uses smooth particle
hydrodynamics (SPH)~\cite{lucy77:sph,gingold77:sph};
see Theuns et al.~\shortcite{theuns99:cosmology,theuns98:apmsph} for
details. These 
simulations use $64^3$ particles of each 
species, so the SPH particle masses are $1.65\times
10^6\,(\Omega_bh^2/0.0125) (h/0.5)^{-3} {\rm M}_\odot$ and the CDM
particles are more 
massive by a factor $\Omega_{\rm CDM}/\Omega_b$. This resolution is
sufficient to simulate line widths reliably
\cite[note that in the hotter
simulations numerical convergence will be even better than in the cooler
model~S, which was investigated in detail by Theuns et
al.~\shortcite{theuns98:apmsph}]{theuns98:apmsph,bryan98:lya_numeffects}. 

We assume that the IGM is ionized and photoheated by an imposed uniform
background of UV-photons that originates from quasars, as computed by
Haardt \& Madau~\shortcite{haardt96:hm_spectrum}. This flux is
redshift dependent, due to the evolution of the quasar luminosity
function. The amplitude of the
flux is indicated as `HM' in the $\Gamma_{\HI}$ column of
Table~\ref{tbl:models} (where $\Gamma_{\HI}$ is the \HI\ ionization
rate due to the ionizing background). For the low $\Omega_bh^2$ models,
we have divided the ionizing flux by two, indicated as `HM/2'. We do
not impose thermal equilibrium but 
solve the rate equations to track the abundances of \HI, \HII\ and
\HeI, \HeII\ and \HeIII. We assume a helium abundance of $Y=0.24$ by
mass. See Theuns et al.~\shortcite{theuns98:apmsph} for further details.

At several output times we compute simulated spectra along 1200 random
lines of sight through the simulation box. Each spectrum is convolved with a
Gaussian with full width at half maximum of FWHM = 8~$\kms$, then
resampled onto pixels of width 3~$\kms$ to mimic the instrumental
profile and characteristics of the HIRES spectrograph on the Keck
telescope. We rescale the background flux in the analysis stage such
that the mean effective optical depth at a given redshift in all models
is the same as for the Ob model. This model has a mean
absorption in good agreement with observations
\cite{rauch97:lya_opacity} (Ob has $\bar\tau_{\rm 
eff}=0.93$, 0.33 and 0.14 at $z=4$, 3 and 2). Finally, we add to the
flux in every pixel a Gaussian random signal with zero mean and
standard deviation $\sigma=0.02$ to mimic noise. The spectra cover a
small enough velocity range to be fit by a flat continuum, as chosen
by a simple procedure~\cite{theuns98:apmsph} described as follows. A
low average continuum is 
assumed initially, then all pixels below and not within 1~$\sigma$ of
this level are rejected and a new average flux level for the remaining
pixels is computed. The last two steps are repeated until the average
flux varies by less than 1\%. This final average flux level is adopted
as the fitted continuum and the spectrum is renormalised accordingly. 
The absorption features in these mock observations are then fitted with
Voigt profiles using an automated version of
\vpfit~\cite{carswell87:vpfit}.

Although the
simulated models have different equations of state, they cover only a
limited part of the possible parameter space $(T_0,\gamma)$. The
models were originally
intended to investigate the dependence of QSO absorption line
statistics on cosmology~\cite{theuns99:cosmology}. Changing the
reionization history can lead to very different values of $T_0$ and
$\gamma$. To quantify the 
relation between the cutoff in the $b(N)$-distribution and the
equation of state, it is necessary to include models covering a wide
range of $T_0$ and $\gamma$. We therefore created models with
particular values of $T_0$ and $\gamma$ by imposing an equation of
state on model~Ob. This was done by moving the SPH~particles in the
temperature-density plane. The new models have the same
three components (low- and high-density power-law equations of state
and shocked gas) as the original model, but a different
equation of state for the low-density gas. In particular, the
intrinsic scatter around the power-law is left unchanged. 

\section{THE $\bmath{\lowercase{b}(N)}$-CUTOFF AND THE
EQUATION OF STATE}

\label{sec:relation}

\begin{figure}
\resizebox{\colwidth}{!}{\includegraphics{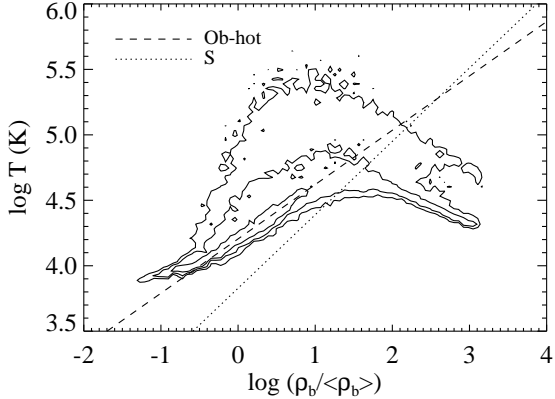}}
\caption{The mass-weighted distribution of fluid elements (SPH
particles) in the temperature-density plane 
for model Ob-hot at $z=3$. Three contour levels are plotted. The
number density of fluid elements increases by an order of magnitude
with each contour level. The dashed line
is a power-law (least absolute deviation) fit to the
temperature-density relation of the gas with density in 
the range $-0.7 \leq \log(\rho_b/\bar{\rho}_b)
\leq 0.3$. Clearly, the vast majority of the low-density
gas follows this tight temperature-density relation. The dotted line
shows the result of a similar fit for model S.} 
\label{fig:tempdens}
\end{figure}

Fig.~\ref{fig:tempdens} shows a contour plot of the mass-weighted
distribution of  fluid elements (SPH particles) from a numerical
simulation in the 
temperature-density diagram. Noting that the number density of fluid elements
increases by an order of magnitude with each contour level, it is
clear that the vast majority of the low-density gas
($\rho_b/\bar{\rho}_b \la 10$) follows a power-law equation of state
(dashed line). The two other components visible in
Fig.~\ref{fig:tempdens} are hot, shocked gas which cannot cool within a
Hubble time and colder, high-density gas for which \HI\ and \HeII\ line
cooling is effective.

\begin{figure*}
\resizebox{\textwidth}{!}{\includegraphics{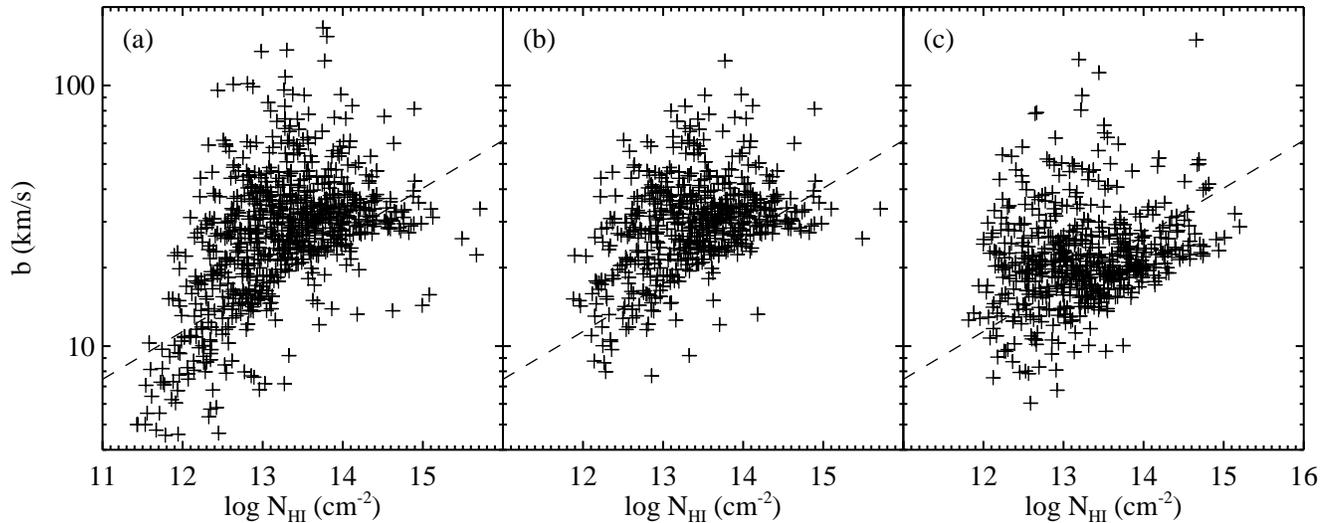}}
\caption{The $b(N)$-distribution for 800 random lines from model
Ob-hot (a,b) and S (c), at $z=3$. The position of each line is
indicated by a cross. Errors are not displayed. In panels (b) and (c)
only those lines are plotted for which  \vpfit\ gives formal errors
$\Delta b/b < 0.5$, $\Delta N/N < 0.5$. The dashed line is the cutoff
for the lines plotted in panel (b) over the range $10^{12.5}\,\cm^{-2}
\le N \le 10^{14.5}\,\cm^{-2}$. Removing the lines with the largest
relative errors in the 
Voigt profile parameters sharpens the cutoff. The $b(N)$-distribution
of the colder model~S clearly cuts off at lower \bpar s than the
$b(N)$-distribution of model~Ob-hot.}  
\label{fig:bN}
\end{figure*}

In Fig.~\ref{fig:bN}a we plot the $b(N)$-distribution for 800
random absorption lines taken from the spectra of model Ob-hot at
redshift $z=3$. A cutoff at low $b$-values, which increases with
column density, can clearly be seen. In Fig.~\ref{fig:bN}b only
those lines for which \vpfit\ gives formal errors in both $b$ and $N$
that are smaller than 50~per~cent are plotted. This excludes most of the lines
that have column density $N\la 10^{12.5}\,\cm^{-2}$ as well
as many of the very narrow lines below the cutoff. Although the
formal errors of the Voigt profile fit given by \vpfit\ have only limited
physical significance, lines in blends tend to have large
errors. Since the $b$-parameters for blended lines can have values
smaller than the minimum set by the thermal smoothing scale (i.e.\
thermal broadening and baryon smoothing), these lines will tend to
smooth out any intrinsic cutoff. Removing the lines with the
largest relative errors therefore results in a sharper cutoff. A smaller
maximum allowed error would result in the removal of many of the
regular, isolated lines.

The large number of data points plotted in Fig.~\ref{fig:bN}a
excludes the possibility that the slope in the cutoff is due to the
large decrease in the number of lines with column density or the
increase in scatter with decreasing column density (Hu et
al.~\shortcite{hu95:lyaobs} reached this conclusion from analysing
spectra that had about 250 absorption lines each). 

The $b(N)$-distribution for the colder model~S is plotted in
Fig.~\ref{fig:bN}c. Clearly, the distribution cuts off at lower
$b$-values. Let us assume that the absence of lines with low
$b$-values is due to 
the fact that there is a minimum line width set by the thermal state
of the gas through the thermal broadening and/or baryon smoothing
scales. Since the temperature of the low-density gas responsible for
the \lya\ forest increases with density (Fig.~\ref{fig:tempdens}), we 
expect the minimum $b$-value to increase with column density,
provided that the column density correlates with the density of the
absorber. 

To see whether this picture is correct, we need to investigate the
relation between the Voigt profile parameters $N$ and $b$, and the
density and temperature of the absorbing gas respectively. Peculiar
velocities and thermal broadening make it difficult to identify the
gas contributing to the optical depth at a particular point in
redshift space. Furthermore, the centre of the Voigt profile fit to an
absorption line will often be offset from the point where the optical
depth is maximum. We therefore need to define a temperature and a
density which are smooth and take redshift space distortions into
account. We choose to use optical depth weighted quantities: the
density of a pixel in velocity space is the sum, weighted by optical
depth, of the density at all the pixels in real space that contribute
to the optical depth of that pixel in velocity space.  We then define
the density corresponding to an absorption line to be the optical
depth weighted density at the line centre. The temperature
corresponding to absorption lines is defined similarly.

\begin{figure*}
\resizebox{\textwidth}{!}{\includegraphics{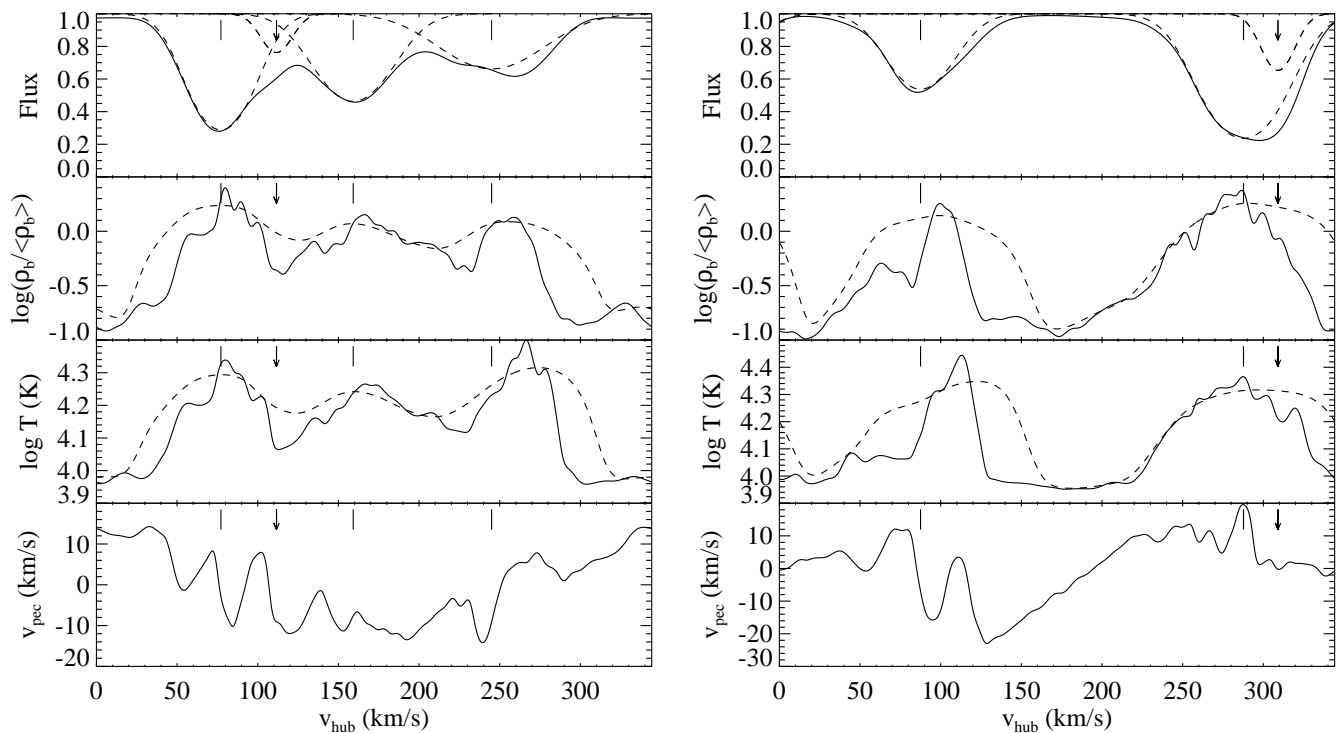}}
\caption{Two random lines of sight through the box of model Ob-hot at
$z=3$. From top to bottom the solid curves are the flux, gas density,
temperature and peculiar velocity respectively, all as a function of
Hubble velocity.
The dashed curves in the top panel are the Voigt profiles
fitted by \vpfit, vertical lines indicate the line centres. The arrows
point at the line centres of lines with a \bpar\
smaller than the cutoff value of the Hui-Rutledge formula
(equation~\ref{eq:hui-rutledge}). These lines
have column densities $N=10^{12.7}\,\cm^{-2}$ (left panel) and
$N=10^{12.9}\,\cm^{-2}$ (right panel), and \bpar s 13.1~$\kms$ (left
panel) and 13.6~$\kms$ (right panel).
They appear as diamonds in Fig.~\ref{fig:cutoff}. The dashed curves
in the second and third panels are the optical depth weighted density
and temperature respectively.}
\label{fig:spectra}
\end{figure*}

In Fig.~\ref{fig:spectra} the optical depth weighted density and
temperature are plotted for two random lines of sight (dashed lines in
the middle two panels), as well as the flux (without noise), and real
space density, temperature and peculiar velocity. The dashed curves in
the top panels are the Voigt profiles fitted by \vpfit, vertical lines
indicate the line centres. Absorption lines correspond to peaks in the
(optically depth weighted) density and temperature, which are strongly
correlated. Although the blends indicated by arrows can be traced back
to substructure in the peaks, their profiles are mainly determined by
the density and temperature of the gas in the main peaks.

\begin{figure}
\resizebox{\colwidth}{!}{\includegraphics{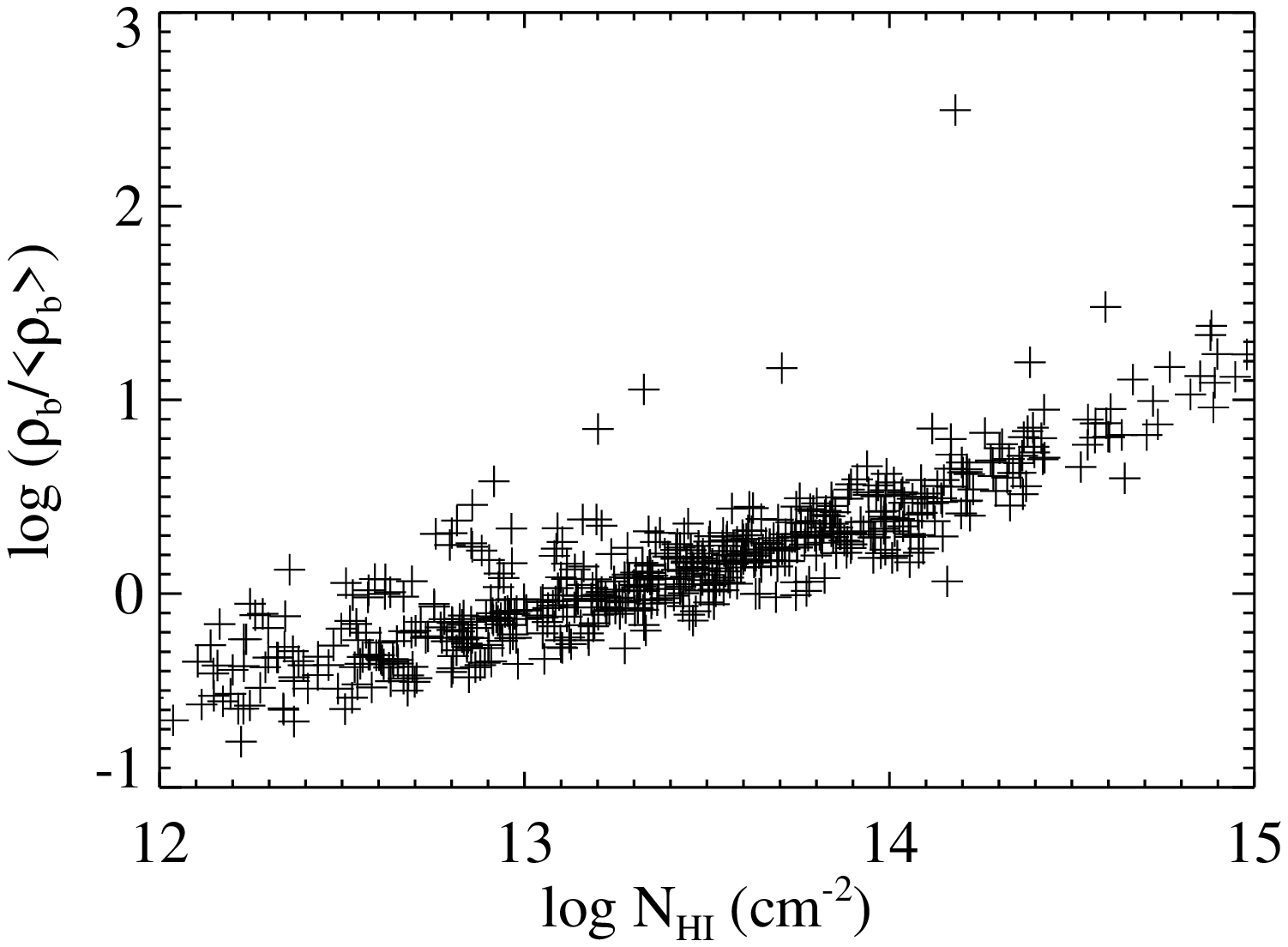}}
\caption{The optical depth weighted gas density at the line centres as a
function of neutral hydrogen column density. Data points correspond to
the lines plotted in Fig.~\ref{fig:bN}b. The column density of a Voigt
profile fit to an absorption line is strongly correlated with the
density of the absorbing gas. The outlier at
$\log(\rho/\bar{\rho_b}) = 2.50$ corresponds to a line in the wing of a
highly saturated line.}
\label{fig:dens-N}
\end{figure}

\begin{figure*}
\resizebox{\textwidth}{!}{\includegraphics{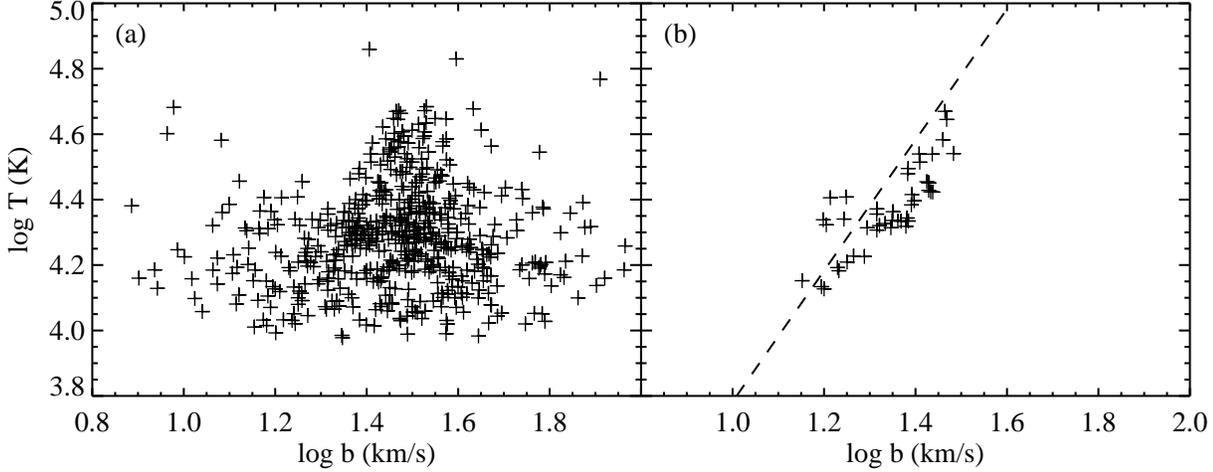}}
\caption{The optical depth weighted temperature at the line centres as
a function of \bpar. The left panel contains all lines plotted in
Fig.~\ref{fig:bN}b. In the right panel only those lines are 
plotted which have \bpar s within one mean absolute deviation of the
power-law fit to the $b(N)$-cutoff plotted as the solid line in
Fig.~\ref{fig:cutoff}. Only for the lines near the cutoff are the
\bpar s correlated with the temperature of the absorbing gas.
The dashed line corresponds to the thermal width,
$b=(2 k_B T/m_p)^{1/2}$.}
\label{fig:temp-b}
\end{figure*}

In Fig.~\ref{fig:dens-N} the optical depth weighted gas density is
plotted as a function of 
column density for the absorption lines plotted in
Fig.~\ref{fig:bN}b. There exists a tight correlation between these two
quantities. Note that lines with column densities $\la
10^{13}\,\cm^{-2}$ correspond to local maxima in underdense regions.
The optical depth weighted temperature is plotted against the
$b$-parameter in Fig.~\ref{fig:temp-b}a. The result is a scatter plot
with no apparent correlation. This is not surprising since many
absorbers will be intrinsically broader than the local thermal
broadening scale. In order to test whether the cutoff in the
$b(N)$-distribution is a consequence of the existence of a minimum
line width set by the thermal state of the gas, we need to look for a
correlation between the temperature and \bpar s of the lines near the
cutoff.

\begin{figure}
\resizebox{\colwidth}{!}{\includegraphics{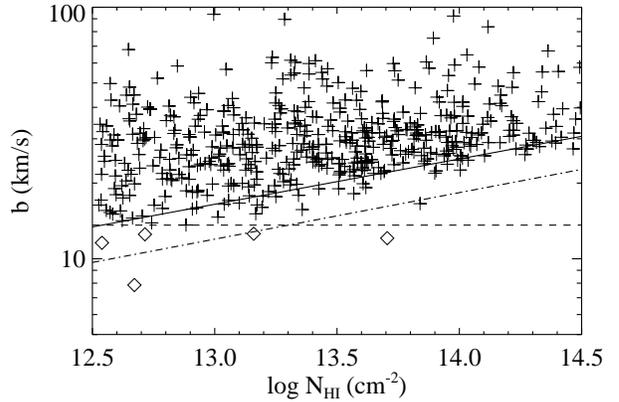}}
\caption{The $b(N)$-distribution for those lines from
Fig.~\ref{fig:bN}b that have column densities in the range
$10^{12.5}\,\cm^{-2} \leq N \leq 10^{14.5}\,\cm^{-2}$. The solid line
is the fitted cutoff 
for this set of lines, the dot-dashed line is the cutoff for the
colder model~S. The horizontal dashed line indicates the minimum $b$-value of
lines used in the fit, lines with smaller \bpar s are indicated
by diamonds. This minimum $b$-value corresponds to the cutoff of the
best-fitting Hui-Rutledge function (see section~\ref{sec:algorithm}).}  
\label{fig:cutoff}
\end{figure}

Determining a cutoff in an objective manner is nontrivial
because of the existence of unphysically narrow lines in blends. We
developed a fitting algorithm that is insensitive to these
lines. This algorithm is described in the next section.
Fig.~\ref{fig:cutoff} is a scatter plot of the lines with column
density in the range $10^{12.5}\,\cm^{-2} \le N \le
10^{14.5}\,\cm^{-2}$. The cutoff fitted to this distribution is also
shown (solid line). The lines that are used in the
final iteration of the fitting algorithm, i.e.\ lines that are close
to the solid line in Fig.~\ref{fig:cutoff}, do indeed display a tight
correlation between the
temperature and $b$-parameter (Fig.~\ref{fig:temp-b}b). The dashed
line in Fig.~\ref{fig:temp-b}b corresponds to the thermal width,
$b=(2 k_B T/m_p)^{1/2}$, where $m_p$ is the mass of a proton and $k_B$ 
is the Boltzmann constant. Lines corresponding to density peaks
whose width in velocity space is much smaller than the thermal
width, have Voigt profiles with this \bpar . Since the temperature
plotted in Fig.~\ref{fig:temp-b} is the smooth, optical depth weighted
temperature, we do not expect the relation between $T$ and $b$ to be
identical to the one indicated by the dashed line, even if all the
line widths were purely thermal.
Although other definitions of the density and temperature are possible
and will give slightly different results, qualitatively the results will
be the same for any sensible definition of these physical quantities.
Figs.\ \ref{fig:dens-N} and \ref{fig:temp-b}b therefore
suggest that the cutoff in the $b(N)$-distribution should be strongly
correlated with the equation of state of the absorbing gas.

Let us look in more detail at the relation between the $b(N)$-cutoff
and the equation of state. We have already shown (Fig.~\ref{fig:tempdens})
that almost 
all the low-density gas follows a power-law equation of state:
\begin{equation}
\log(T) = \log(T_0) + (\gamma-1)\log(\rho/\bar{\rho}).
\label{eq:eos}
\end{equation}
The relations between the density/temperature and column density/\bpar\
of the absorption lines near the cutoff can also be fitted by
power-laws (Fig.~\ref{fig:dens-N} and Fig.~\ref{fig:temp-b}b):
\begin{equation}
\log(\rho/\bar{\rho}) = A + B\log(N/N_0),
\label{eq:rho-N}
\end{equation}
\begin{equation}
\log(T) = C + D\log(b).
\label{eq:T-b}
\end{equation}
Combining these equations, we find that the $b(N)$-cutoff is also a
power-law,
\begin{equation}
\log(b) = \log(b_0) + (\Gamma-1)\log(N/N_0),
\label{eq:b-N}
\end{equation}
whose coefficients are given by,
\begin{equation}
\log(b_0) = {1 \over D} \left [ \log(T_0) - C + (\gamma-1)A \right ],
\label{eq:logb_0}
\end{equation}
\begin{equation}
\Gamma-1 = {B \over D}(\gamma-1).
\label{eq:Gamma}
\end{equation}
Hence the intercept of the cutoff, $\log b_0$, depends on both the
amplitude and the slope of the equation of state, while the slope of
the cutoff, $\Gamma-1$, is proportional to the slope of the
equation of state. If we set $N_0$ equal to the column density
corresponding to the mean gas density, then the coefficient $A$ vanishes
and $\log b_0$ no longer depends on $\gamma$.

The cutoff in the $b(N)$-distribution is measured over a certain
column density range. We choose to measure the cutoff over the
interval $10^{12.5}\,\cm^{-2} \le N \le 10^{14.5}\,\cm^{-2}$. This interval
corresponds roughly to the gas density range for which the equation of
state is well fitted by a power-law. For lines with column density
$N\la 10^{12.5}\,\cm^{-2}$, the scatter in the observations becomes
very large due to 
noise and line blending. The relation between the gas overdensity and
\HI\ column density depends on redshift. Hence different column density
intervals should be used for different redshifts if one wants to
compare the equation of state for the same density range.

Fitting a cutoff to a finite number of lines introduces statistical
uncertainty in the measured coefficients. We minimize the correlation
between the errors in the 
coefficients by subtracting the average abscissa value (i.e.\
$\log N$) before fitting the cutoff. This can be done by setting
$\log N_0$ in equation~\ref{eq:b-N} equal to the mean $\log N$ of the
lines in the column density range over which the cutoff is
measured. This column density does in general not correspond to the
mean gas density and hence $\log b_0$ will in general depend on
$\gamma$. We will show below that this dependence can be removed by
renormalising the equation of state.

\begin{figure}
\resizebox{\colwidth}{!}{\includegraphics{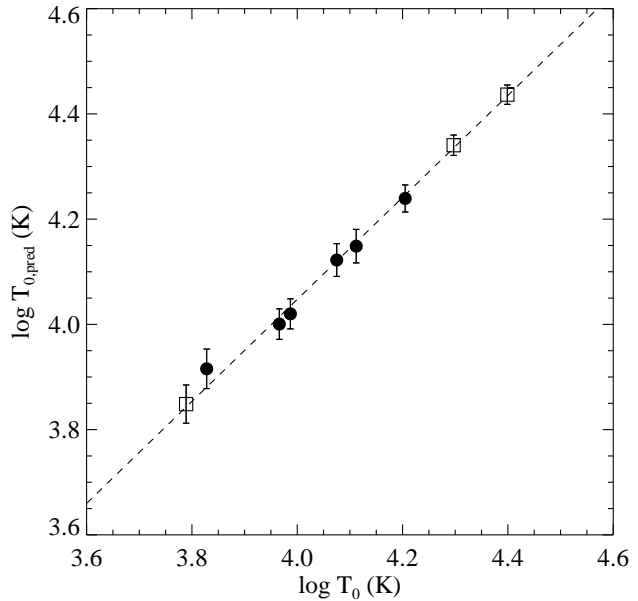}}
\caption{The temperature at mean density predicted from the power-law model
(equation~\ref{eq:logb_0}) as a function
of the true temperature at redshift $z=3$. Error bars
enclose 68\%
confidence intervals around the medians, as determined from 500
sets of 300 random lines. Open squares indicate that the
points resulted from imposing an equation of state with the given $T_0$
on model~Ob. All points have similar values of $\gamma$. The dashed
line is the least-squares fit.}
\label{fig:pred_temp}
\end{figure}

\begin{figure}
\resizebox{\colwidth}{!}{\includegraphics{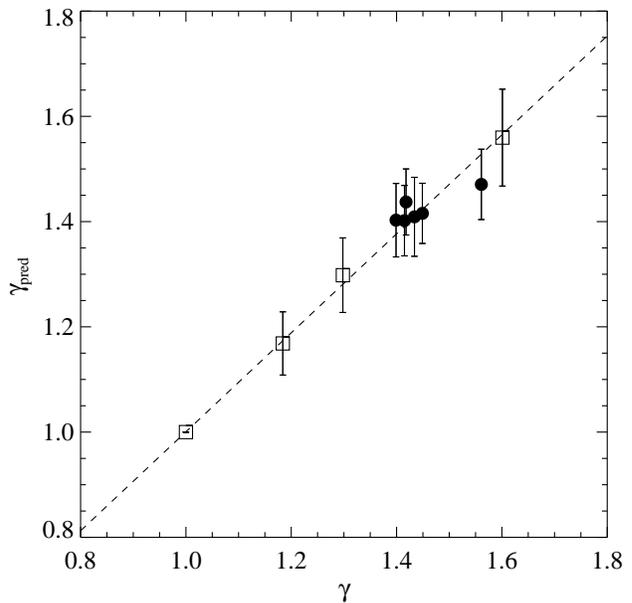}}
\caption{The index $\gamma$ predicted from the power-law model
(equation~\ref{eq:Gamma}) as a function 
of the true index of the power-law equation of state at redshift
$z=3$. Error bars enclose 68\%
confidence intervals around the medians, as determined from 500
sets of 300 random lines. Open squares indicate that the
points resulted from imposing an equation of state with the given $\gamma$
on model~Ob, while leaving $T_0$ unchanged. The dashed
line is the least-squares fit.}
\label{fig:pred_gamma}
\end{figure}

In Fig.~\ref{fig:pred_temp} we plot the temperature at mean density
predicted from the power-law model (equation~\ref{eq:logb_0}) as a
function of the true $T_0$. Data points are determined by fitting
power-laws to 500 sets of 
300 random absorption lines. Error bars enclose 68~per~cent confidence
intervals around the medians. Solid circles are used for data from
simulated models, open squares are used for models created by imposing
an equation of state on model~Ob. These conventions will be used
throughout the paper. Fig.~\ref{fig:pred_gamma} is a similar plot for
the slope of the equation of state (equation~\ref{eq:Gamma}). The
predicted and true parameters of the equation of state are 
highly correlated. The slight offset between the
predicted and true quantities simply reflects the fact that the
optical depth weighted density and temperature are not exactly the
same as the true density and temperature of the absorbing gas (i.e.\
the $T$ and $\rho$ appearing in equation~\ref{eq:eos} are not exactly
the same as those appearing in equations \ref{eq:rho-N} and
\ref{eq:T-b}). The main conclusion to draw from these plots is that
the power-law model works and that we can therefore use these
equations to gain insight in the relationship between the equation of 
state and the cutoff in the $b(N)$-distribution. 

\begin{figure}
\resizebox{\colwidth}{!}{\includegraphics{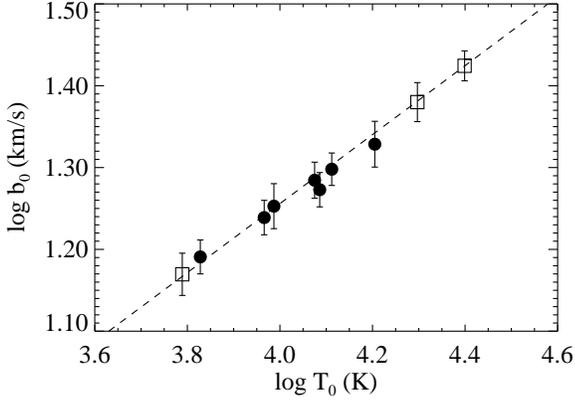}}
\caption{The intercept of the $b(N)$-cutoff as a function of the temperature
at mean density at redshift $z=3$. The error bars enclose 68\%
confidence intervals around the medians, as determined from 500
sets of 300 random lines. Open squares indicate that the
points resulted from imposing an equation of state with the given $T_0$
on model Ob. All points have similar values of $\gamma$. The dashed
line is the maximum likelihood fit, as determined from the
full probability distributions. The tight correlation and the
relatively small errors make the intercept of the cutoff a sensitive
measure of the amplitude of the equation of state.}
\label{fig:intercept-T}
\end{figure}

The objective is to establish the relations between
the cutoff parameters and the equation of state using simulations. These
relations can then be used to measure the equation of state of the IGM using
the observed cutoff in the $b(N)$-distribution. 
The amplitudes of the power-law fits to the cutoff and the equation of
state are plotted against each other in
Fig.~\ref{fig:intercept-T}. The relation between $\log b_0$ and $\log
T_0$ is linear, implying that the coefficients appearing in
equation~\ref{eq:logb_0} do not vary strongly with cosmology. 
The error bars, which indicate the dispersion in
the cutoff of sets of 300 lines (typical for $z=3$), are small compared
to the differences between the models. This means that measuring the
cutoff in a single QSO spectrum can provide significant constraints on
theoretical models.

\begin{figure}
\resizebox{\colwidth}{!}{\includegraphics{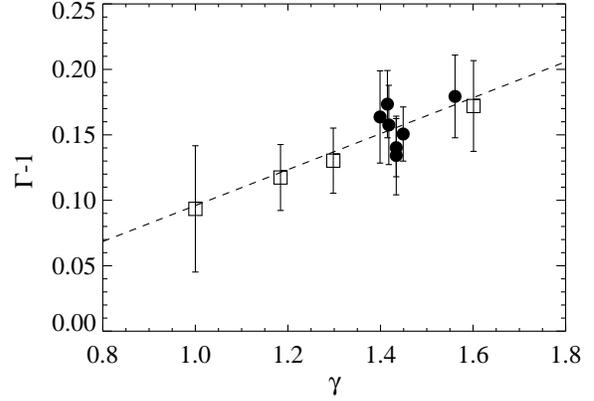}}
\caption{The slope of the $b(N)$-cutoff as a function of the index of the
power-law equation of state for redshift $z=3$. The error bars enclose 68\%
confidence intervals around the medians, as determined from 500
sets of 300 random lines. Open squares indicate that the
points resulted from imposing an equation of state with the given $\gamma$
on model Ob, while leaving $T_0$ unchanged. The dashed
line is the maximum likelihood fit, as determined from the
full probability distributions. The slope of the cutoff is a measure
of the slope of the equation of state, The weak dependence of $\Gamma$
on $\gamma$ and the large variance in the measured $\Gamma$ will make
it hard to constrain $\gamma$ accurately.}
\label{fig:slope-gamma}
\end{figure}

The slope of the cutoff, $\Gamma-1$, is plotted against $\gamma$ in
Fig.~\ref{fig:slope-gamma}. The relation between the two is linear,
but $\Gamma$ increases only slowly with $\gamma$. The dispersion in
the slope of the cutoff for a fixed equation of state is 
comparable to the difference between the models\footnote{The simulated
models all have similar values of $\gamma$ because they all have the
same $UV$-background.}. The weak dependence of $\Gamma$ on $\gamma$
and the large spread in the measured $\Gamma$ will make it difficult
to put tight constraints on the slope of the equation of state.

\subsection{Correlations}

\begin{figure*}
\resizebox{\textwidth}{!}{\includegraphics{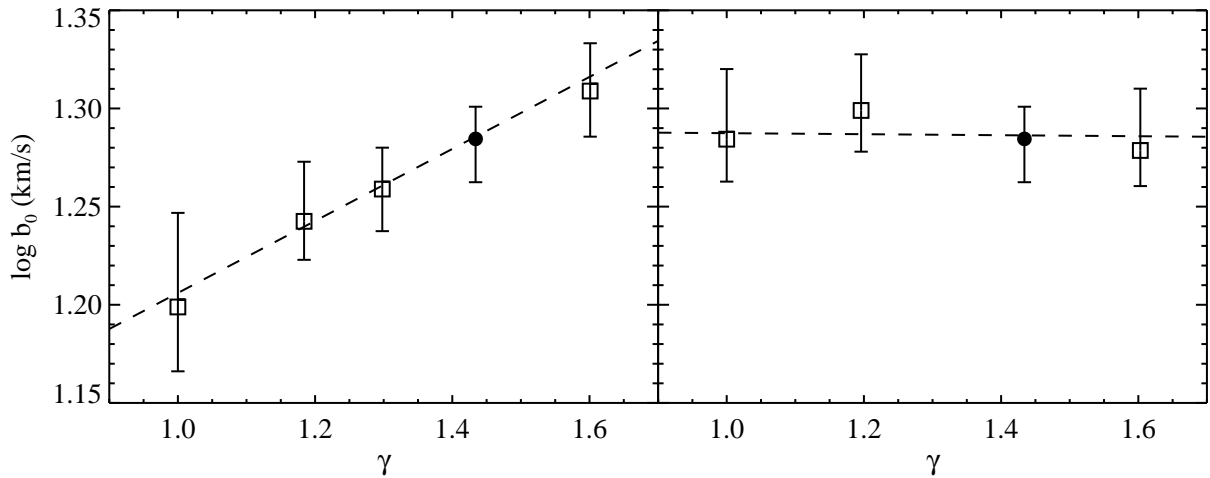}}
\caption{The intercept of the $b(N)$-cutoff as a function of the index
$\gamma$. The models in the left panel all have the same $T_0$, while
the models in the right panel have identical $T_\delta$, where
$\delta=2$. The intercept of the cutoff is a measure of the temperature of
gas at the characteristic density corresponding to the absorption
lines used to fit the cutoff (which is $\delta \approx 2$ for the
interval $10^{12.5}\,\cm^{-2} \le N \le 10^{14.5}\,\cm^{-2}$ at
$z=3$). Hence the intercept 
depends on the slope of the equation of state, unless it is normalised to the
temperature at this density. Plotted data points are Ob (filled
circle) and different equations of state imposed on Ob (open squares),
all for 300 random lines at $z=3$. Error bars enclose 68\%
confidence intervals around the median values. The dashed
lines are the maximum likelihood fits.} 
\label{fig:gammacorr}
\end{figure*}

The intercept of the
cutoff in the $b(N)$-distribution is a measure of the temperature at
the characteristic density of the absorbers corresponding to the lines
used to fit the cutoff. In general this is not the mean
density and consequently the translation from the intercept $b_0$, to
the temperature at mean density $T_0$, depends on the slope
$\gamma$. This is illustrated in the left panel of
Fig.~\ref{fig:gammacorr}, where the measured values of $b_0$ for models that
have identical values of $T_0$, but a range of $\gamma$-values are
compared. As predicted by the power-law model
(equation~\ref{eq:logb_0}), $\log b_0$ increases linearly with $\gamma$.

In principle, the index $\gamma$ can be
measured using the slope of the cutoff, to which it is proportional
(equation~\ref{eq:Gamma}). However, $\Gamma$ increases only slowly
with $\gamma$ and the statistical variance in $\Gamma$ is large,
making it hard to put tight constraints on $\gamma$. It appears
therefore that even though $\log b_0$ is very sensitive to $\log T_0$,
and can be measured very precisely, the uncertainty in $T_0$ will be
relatively large due to the weak constraints on $\gamma$.
It is important to realise that any statistic that is sensitive to the
temperature, depends in general on both $T_0$ and $\gamma$. It is for
example impossible to determine $T_0$ by fitting the \bpar\
distribution at $N \neq N(\bar{\rho})$. Since this statistic is
sensitive to the temperature at the density corresponding to a column
density $N$, any equation of state $(T_0,\gamma)$
that has the correct temperature at this density will fit the data.

Although it is conventional to normalise the equation of
state to the temperature at the mean density, it can in principle be
normalised at any density, $\rho_\delta \equiv \bar{\rho}(1+\delta)$ say,
\begin{equation} 
\log T=\log T_\delta + (\gamma-1)\log(\rho/\rho_\delta).
\end{equation}
Similarly, equations~\ref{eq:rho-N} and \ref{eq:b-N} can be generalised to
\begin{equation}
\log (\rho/\rho_\delta) = B\log(N/N_\delta),
\end{equation}
\begin{equation}
\log b = \log b_0 + (\Gamma -1)\log (N/N_\delta),
\end{equation}
where $N_\delta \equiv N(\rho=\rho_\delta)$ and consequently the
coefficient $A$ vanishes. In practice, the cutoff is fitted over a
given column density interval and $\log N_\delta$ is set equal to the mean
$\log N$ of lines in this interval. The measured cutoff is then
converted into an equation of state normalised at the corresponding
density, $\rho_\delta =
\rho(\log N_\delta=\left <\log N \right >)$. Equation~\ref{eq:logb_0}
then becomes $\log b_0 = (\log T_\delta -C)/D$ and the intercept of
the cutoff depends only on the amplitude of the equation of state.

At redshift $z=3$, using the column density
interval $10^{12.5}\,\cm^{-2} \le N \le 10^{14.5}\,\cm^{-2}$,
$N_\delta=10^{13.6}\,\cm^{-2}$  and $\delta 
\approx 2$. In the right panel of Fig.~\ref{fig:gammacorr} the
intercept of the cutoff is plotted as a function of $\gamma$ for a set
of models that all have the same temperature at this density. As
expected, the intercept is insensitive to the slope of the equation of
state.

In summary, the intercept of the $b(N)$-cutoff is a measure of the
temperature of the gas responsible for the absorption lines that are
used to determine the cutoff. If we normalise the equation of
state to the temperature at the characteristic density of the gas,
then the intercept of the cutoff depends only on the amplitude of the
equation of state. The slope of the cutoff is always determined by the slope
of the equation of state. Hence the cutoff in the $b(N)$-distribution
can be used to determine both $T_\delta$, where $\delta$ is
the density contrast corresponding to the mean $\log N$ of the
lines used in the fit, and $\gamma$. The temperature at mean density,
$T_0$, depends on both $T_\delta$ and $\gamma$ and therefore on both the
intercept and the slope of the cutoff.

\section{MEASURING THE CUTOFF}

\label{sec:algorithm}

The main problem in measuring the cutoff in the $b(N)$-distribution is
the fact that it is contaminated by spurious narrow lines. Line blending and
blanketing, noise and the presence of unidentified metal lines all
give rise to absorption lines that are narrower than the lower limit
to the line width set by the thermal state of the gas. We have
therefore developed an iterative procedure for fitting the cutoff
that is insensitive to the presence of a small number of 
narrow lines.  In order to minimize the effects 
of outliers, robust least absolute deviation fits are used. The first
step is to fit a 
power-law to the entire set of lines. 
Then the lines that have \bpar s more than one mean absolute deviation
above the fit are removed and a power-law is fitted to the remaining
lines. These last two steps are repeated until convergence is
achieved. Finally, the lines more than one mean absolute deviation
below the fit are also taken out and the fit to the remaining lines is
the measured cutoff. The
algorithm works very well if there 
are not too many unphysically narrow lines. At the lowest column
densities ($\la 10^{12.5}\,\cm^{-2}$) however, blends dominate and the
cutoff is washed out. 

Fortunately,
there are ways to take out most of these unphysically narrow lines. 
We have already shown (Fig.~\ref{fig:bN}) that removing lines with
large relative errors in the Voigt profile parameters
significantly sharpens the cutoff. We choose to consider only those
lines with relative errors smaller than 50 per cent. A smaller
maximum allowed error would result in the removal of many of the
regular, isolated lines.

\begin{figure}
\resizebox{\colwidth}{!}{\includegraphics{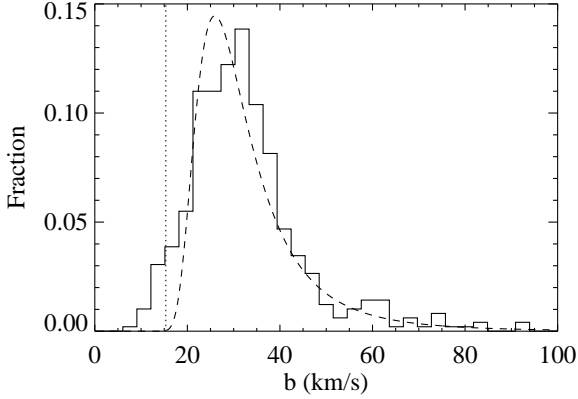}}
\caption{The \bpar\ distribution of the lines plotted in
Fig.~\ref{fig:cutoff}. The dashed curve is the best-fitting Hui-Rutledge
function (equation~\ref{eq:hui-rutledge}) for this distribution. The
vertical dotted line indicates the point where this function has the
value $10^{-4}$, which corresponds to the horizontal dashed line in
Fig.~\ref{fig:cutoff}. Virtually all of the lines that are narrower
than this cutoff in the Hui-Rutledge function are blends. We remove
these lines from the sample before fitting the $b(N)$-cutoff.}  
\label{fig:bpar}
\end{figure}

Another cut in the set of absorption lines can be made on the basis of
theoretical arguments.
Assuming that absorption lines arise from
peaks in the optical depth $\tau$, and assuming that $\ln\tau$ is a
Gaussian random variable (as is the case for linear fluctuations), Hui \&
Rutledge~\shortcite{hui97:bdistr} derive a single parameter analytical
expression for the $b$-distribution:
\begin{equation}
{dN \over db} \propto {b_\sigma^4 \over b^5} \exp \left [ -{b_\sigma^4
\over b^4} \right ],
\label{eq:hui-rutledge}
\end{equation}
where $b_\sigma$ is determined by the average amplitude of the
fluctuations and by the effective smoothing scale. Fig.~\ref{fig:bpar}
shows the \bpar\ distribution for the lines plotted in
Fig.~\ref{fig:cutoff} and the best-fitting Hui-Rutledge
function (dashed line). The $b$-value for which the Hui-Rutledge fit
vanishes (dotted line) corresponds to the
dashed line in Fig.~\ref{fig:cutoff}. 
The $b$-distribution has a tail of narrow lines which is not present in
the theoretical Hui-Rutledge function. These lines are indicated by
diamonds in Fig.~\ref{fig:cutoff}. Direct inspection shows that all
these lines occur in blends. Two examples are the lines indicated by
arrows in Fig.~\ref{fig:spectra}. The size of the low
$b$-tail depends on the number of blended lines and therefore on the
signal to noise of the spectrum and on the fitting procedure. However,
we find that in general, virtually all of the lines with $b$-values
smaller than the 
cutoff in the fitted Hui-Rutledge function are blends. We therefore
remove these lines before fitting the cutoff in the $b(N)$-distribution.

\begin{figure*}
\resizebox{\textwidth}{!}{\includegraphics{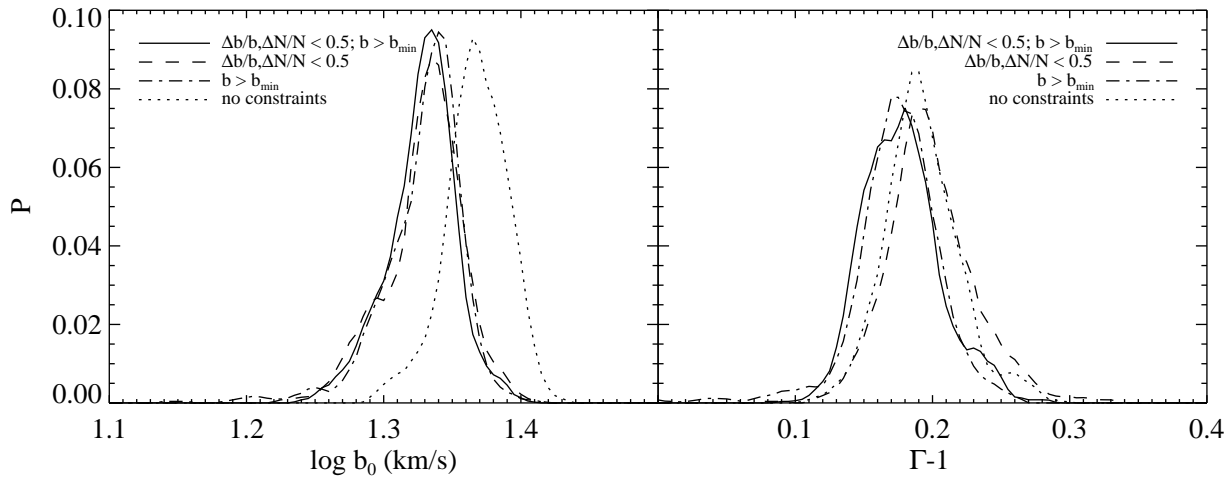}}
\caption{Probability distributions for the intercept (left panel) and
slope (right panel) of the cutoff in the $b(N)$-distribution for 300
random absorption lines from model Ob-hot at $z=3$. The solid line is for
the algorithm used 
throughout this paper. The other curves illustrate the effect of
including lines with \bpars\ smaller than the cutoff of the
Hui-Rutledge function (dashed), relative errors greater than 50\% (dot-dashed)
and both (dotted). All cutoffs are for the column density range
$10^{12.5}\,\cm^{-2} \leq N \leq 10^{14.5}\,\cm^{-2}$.}
\label{fig:cutoff-vars}
\end{figure*}

Fig.~\ref{fig:cutoff-vars} illustrates the effect of the two cuts
(relative errors and Hui-Rutledge function). The probability
distributions for the parameters of the fitted cutoff, $b_0$
and $\Gamma$, are plotted for different cuts. The dotted lines are the
distributions resulting from fitting the cutoff for the complete set
of lines. Removing the lines with
large errors or those with $b$-values smaller than the cutoff of the
Hui-Rutledge fit results in a \emph{smaller} intercept $\log b_0$\footnote{The
removal of very narrow lines reduces the scatter around the cutoff and
therefore causes the algorithm to converge at a lower mean absolute
deviation. More iterations are needed before convergence is obtained,
yielding a lower final cutoff.}. 
It makes no difference which cut is applied. Applying a cut in
error-space does not affect
the slope of the cutoff, $\Gamma-1$. However, taking out the
lines below the Hui-Rutledge cutoff, removes the low-$b$, low-$N$ tail
without affecting the higher column density end, and therefore yields a
smaller slope.

A typical QSO spectrum at $z\sim 3$ has about 300 \lya\ absorption
lines between its \lya\ and \lyb\
emission lines with column densities in the range $10^{12.5}\,\cm^{-2}
\le N \le 10^{14.5}\,\cm^{-2}$. The number density of 
\lya\ lines decreases rapidly with decreasing redshift. At $z\sim 2$
there are typically less than 100 lines. The fact that the number of
lines in an observed \lya\ forest is finite introduces statistical
variance. We therefore use many (500) realizations to determine the full
probability distributions of the parameters of the cutoff.

\begin{figure}
\resizebox{\colwidth}{!}{\includegraphics{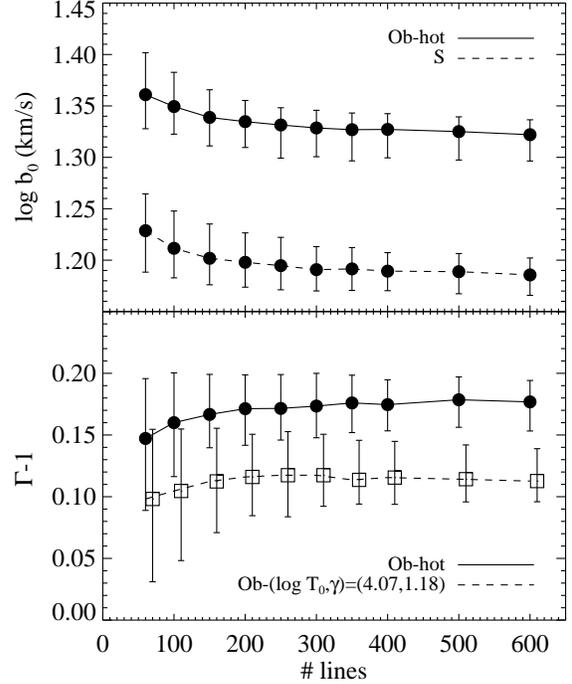}}
\caption{The intercept (upper panel) and slope (bottom panel) of the
$b(N)$-cutoff for different numbers of absorption lines used in the fit. 
The cutoff depends weakly on the number of lines, although the
statistical variance does increase significantly when the number of
lines falls below 200. The numbers refer to the number of lines after
removal of the lines with errors greater than 50\%, but
before applying the Hui-Rutledge cut. The error bars indicate 68\%
confidence intervals around the medians. The lower curve in the bottom
panel has been slightly displaced.}
\label{fig:convergence}
\end{figure}

Fig.~\ref{fig:convergence} illustrates the effects of changing the
number of absorption lines. The algorithm is surprisingly insensitive
to the number of lines. The parameters of the cutoff vary only
slightly for 60 lines or more. The variance does decrease as
the number of lines increases, but remains almost the same for more than
200 lines. This suggests that the method should work even with the
small number of lines per spectrum at $z\sim 2$. It also
opens up the possibility of splitting higher redshift spectra into
redshift bins. The fact that the cutoff depends weakly on the
number of lines is not a problem, since we can determine the relation
between the cutoff and the equation of state for any number of lines,
in particular for the number of lines in an observed spectrum.

While the statistical variance in the
intercept of the $b(N)$-cutoff is small, the variance in the slope is
comparable to the differences between models. Fortunately, we can do
better than measuring the cutoff for the complete set of lines in an
observed spectrum. The bootstrap method (drawing $n$ random lines from the
complete set of $n$~lines, with replacement) can be used to generate a
large number of synthetic data sets. These data sets can then be used
to obtain approximations to the probability distributions for the
parameters of the cutoff. Since bootstrap resampling replaces a random
fraction of the original lines by duplicated original
lines, a smaller fraction of the $b$-$N$ space around the cutoff is filled
and the variance in the measured cutoff increases. Although the bootstrap
distribution is generally broader than the true distribution, its
median is a robust estimate of the true median. When dealing with
observed spectra, we will use the medians
of the bootstrap distributions as our best estimates of the parameters
of the $b(N)$-cutoff. In section~\ref{sec:montecarlo} we will use
Monte Carlo simulations to estimate the variance in the bootstrap
medians.

\section{SYSTEMATIC EFFECTS}

\label{sec:systematics}

In section~\ref{sec:relation} we established the relation between the
cutoff in the $b(N)$-distribution and the equation of state of the
low-density gas. In this section we will investigate whether other
processes can change this relation. 

\subsection{Cosmology}

\begin{figure*}
\resizebox{\textwidth}{!}{\includegraphics{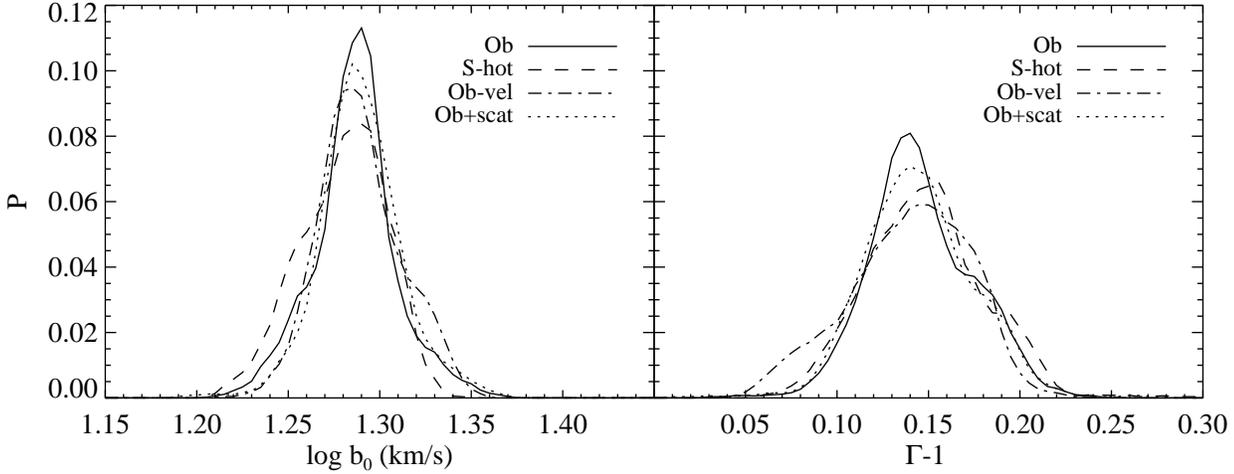}}
\caption{Probability distributions for the intercept (left panel) and
slope (right panel) of the $b(N)$-cutoff for models with identical
equations of state, but different systematics. S-hot (dashed line) is
model~S with the equation of 
state of model~Ob imposed. Ob-vel (dot-dashed) is identical to
model~Ob, but with all peculiar velocities set to zero. Ob+scat is model~Ob
with twice as much scatter around the equation of state.
All distributions are determined
using 500 sets of 300 random absorption lines, at $z=3$.} 
\label{fig:systematics}
\end{figure*}

Cosmology affects not only the equation of state, but
also determines the evolution of structure. Theuns et
al.~\shortcite{theuns99:cosmology} showed that the $b$-parameter
distribution depends on cosmology. In particular, they showed that
the effect of peculiar velocity gradients on the line widths can be
very different in different CDM variants. 
We have recomputed simulated spectra for model~S after imposing the
equation of state of the significantly hotter model~Ob. 
In Fig.~\ref{fig:systematics} the probability distributions of the
parameters of the $b(N)$-cutoff are compared for model~Ob (solid
lines) and this new model, S-hot (dashed).
The distributions are almost indistinguishable. Also
plotted is model~Ob-vel (dot-dashed), which was created by setting all
peculiar velocities in model~Ob to zero. Again, the probability
distributions are almost unchanged. 

\begin{figure*}
\resizebox{\textwidth}{!}{\includegraphics{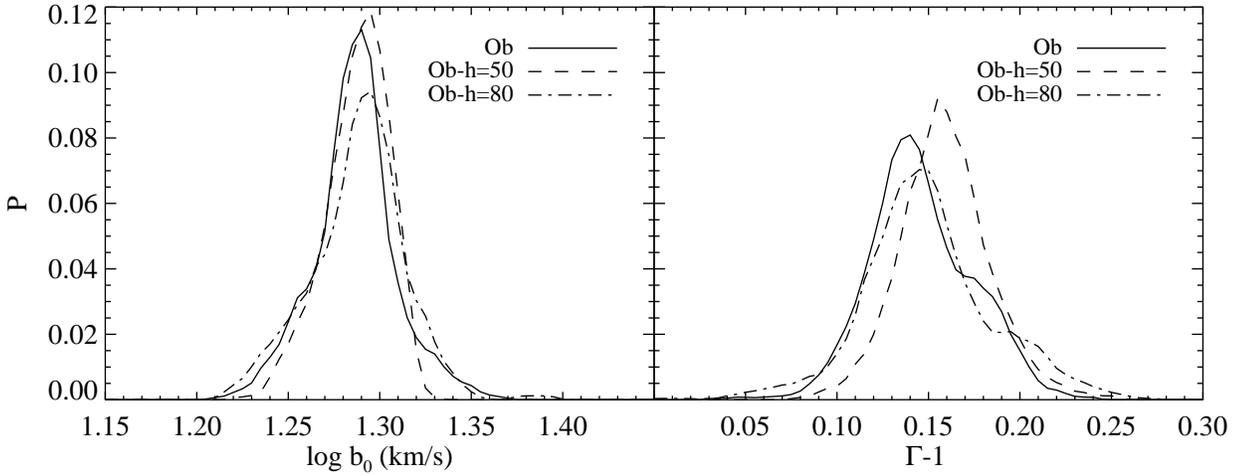}}
\caption{Probability distributions for the intercept (left panel) and
slope (right panel) of the $b(N)$-cutoff for models with identical equations
of state, but different Hubble flow contributions to the line
widths. For models Ob-h=50 and Ob-h=80 the spectra have 
been calculated using a Hubble parameter corresponding to a present
day value of $h=0.5$ and $h=0.8$ respectively (Ob has $h=0.65$). All
distributions are determined using 500 sets of 300 random absorption
lines, at $z=3$.}  
\label{fig:hubble}
\end{figure*}

The line widths of many of the low column density lines are dominated
by the Hubble flow. Fig.~\ref{fig:hubble} illustrates the effect that
changing the Hubble expansion has on the cutoff in the
$b(N)$-distribution. The obvious way to change the Hubble flow is to
change the Hubble constant. However, the equation of state also
depends on the value of the Hubble constant. In order to isolate the
effect of the Hubble flow, we changed the value of the Hubble constant in
the analysis stage, i.e.\ just before computing the
spectra, keeping the equation of state fixed. Increasing the value of
the Hubble parameter at $z=3$ from a 
corresponding present day value of $h=0.65$ to 0.8 has no effect on
the cutoff. Lowering $h$ to 0.5 shifts the slope 
to slightly larger values, but leaves the intercept unchanged.

We conclude that, unlike the $b$-distribution, the cutoff in the
$b(N)$-distribution is independent of the assumed CDM model for a
fixed equation of state.

\subsection{Mean absorption}

The optical depth in neutral hydrogen is proportional to the quantity
$\Omega_b^2h^3/\Gamma_{\HI}$. Since it is still unclear what the
dominant source of the metagalactic ionizing background is, the
\HI\ ionization rate, $\Gamma_{\HI}$, is uncertain.
Changing $\Gamma_{\HI}$ has very little 
effect on the equation of state~\cite{hui97:tempdens}, which means
that the optical depth can be scaled to match the observed mean flux
decrement in the analysis stage. However, the observations do show
some scatter in the mean flux decrement. This scatter could be
due to spatial variations of the ionizing background, or it could be
caused by measurement errors. We therefore need to check whether
errors in the assumed effective optical depth affect the relation
between the cutoff and the equation of state.

\begin{figure}
\resizebox{\colwidth}{!}{\includegraphics{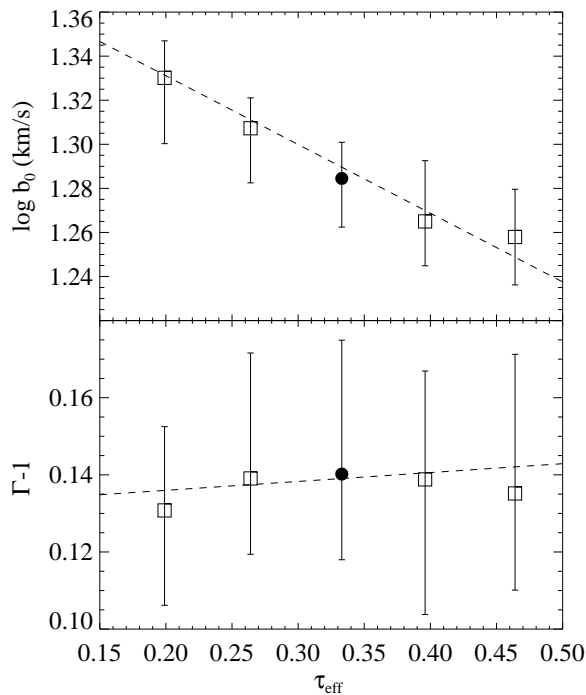}}
\caption{The intercept of the $b(N)$-cutoff decreases slightly with
increasing absorption (upper panel). The slope of the cutoff and the
effective optical depth appear to be uncorrelated (bottom panel). For
realistic errors in the assumed effective optical depth (10--20\% at
$z\sim 3$), the error in the intercept is small compared to the
statistical variance. Plotted
data points are model~Ob (filled circle) and models with rescaled background
fluxes (open squares), all for 300 random
lines at $z=3$. All models have the same equation of state. Error bars
enclose 68\% confidence intervals around the median values.} 
\label{fig:abscorr}
\end{figure}

Changing the effective optical depth by rescaling the ionizing background
alters the relation between column density and gas
density. This will shift the $b(N)$-distribution along the
$N$-axis. Hence we expect the intercept of the cutoff to change and
the slope to remain constant. In terms of our power-law model,
increasing the photoionization rate (i.e.\ decreasing the effective
optical depth) will increase the coefficient $A$
in equation~\ref{eq:rho-N} and thus increase the measured intercept
$b_0$ for a given $T_0$ (equation~\ref{eq:logb_0}), while leaving the
slope $\Gamma-1$ constant
(equation~\ref{eq:Gamma}). Fig.~\ref{fig:abscorr} confirms these  
predictions. The dependence of $b_0$ on the effective optical depth
turns out to be rather weak, even for models with a
relatively steep cutoff. Realistic errors in the mean
flux decrement (10--20\% at $z\sim 3$) will give rise to errors in
$b_0$ that are smaller than the statistical variance. 

\subsection{Signal to noise}

\begin{figure}
\resizebox{\colwidth}{!}{\includegraphics{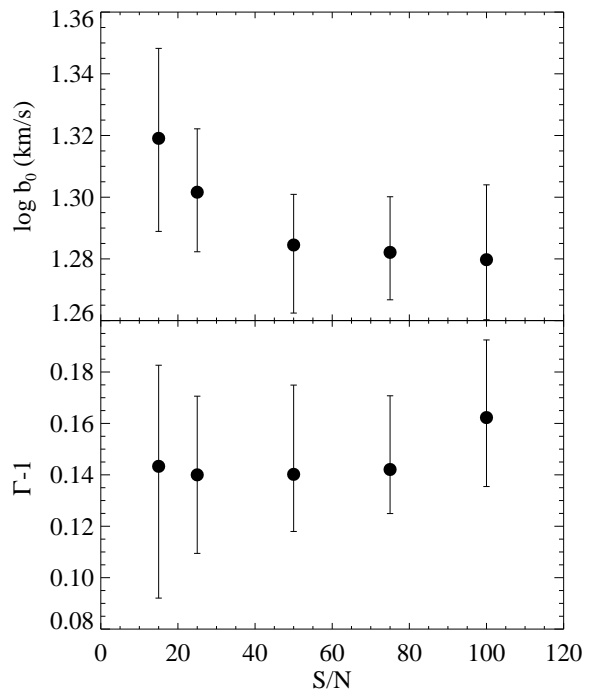}}
\caption{The intercept (top panel) and slope (bottom panel) of the
$b(N)$-cutoff as a function of signal to noise ratio (S/N). Variations in
the S/N are unimportant as long as the S/N is greater than about~20.
Plotted data points are all for 300 random 
lines taken from model~Ob at $z=3$. All models have the same equation
of state. Error bars enclose 68\% 
confidence intervals around the median values.} 
\label{fig:sncorr}
\end{figure}

The signal to noise ratio (S/N) per pixel in the simulated spectra is 50,
comparable to the noise level in high-quality observations. However,
spectra taken with for example the HIRES spectrograph on the KECK
telescope have a S/N that varies across the spectrum.
Fig.~\ref{fig:sncorr} illustrates the effect of changing the signal to
noise (S/N) in the spectrum. The statistical variance in the cutoff
increases rapidly when the S/N falls below 25. While the intercept
increases slightly for S/N smaller than 25, the slope appears to be
independent of the signal to noise. We therefore conclude that
variations in the S/N in observed spectra are unimportant, as long as
the signal to noise is greater than about 20. 

\subsection{Missing physics}

Since the cosmological parameters and reionization history have yet to
be fully constrained, the models used in this paper may not be correct.
This will however not affect the results of this paper, as long as the
bulk of the low-density gas follows a power-law equation of state. We
should therefore ask if there are any physical processes,
which have not been incorporated in the simulations, that could
destroy the uniformity of the thermal state of the low-density
gas. Feedback from massive stars and fluctuations in the spectrum of
the photoionizing flux are two examples. Although these
processes would predominantly affect the gas in virialized halos, they
could introduce some additional scatter in the equation of state of
the gas responsible for the low column density \lya\ forest.

We recomputed simulated spectra for model~Ob, after doubling the
scatter around the fitted temperature-density relation. The
$b(N)$-cutoff of this new model, Ob+scat, is indistinguishable from
the one of model~Ob (solid and dotted curves in
Fig.~\ref{fig:systematics}). Note that although processes like
feedback can heat the IGM locally, it is hard to think of any process,
apart from adiabatic expansion, that could cool the low-density
gas. Furthermore, whereas low-density gas that is 
heated to $T>T_0(\rho/\bar{\rho})^{\gamma-1}$ can only cool over a
Hubble time, gas that is cooled below the equation of state is quickly
reheated. Any additional scatter in the thermal state is thus unlikely
to alter the cutoff in the gas temperature.

Any comparison between observations and simulations is complicated by
the fact that observed spectra cover a much larger redshift path than
the simulation boxes and therefore include effects like redshift
evolution and cosmic variance, which are not present in the
simulations. Large-scale fluctuations, especially in the ionization
fraction (i.e.\ the relation between density and column density) could
potentially distort the relation between the mean equation of state
and the $b(N)$-cutoff.

Redshift evolution mainly affects the mean absorption and does so in a
well defined manner, which can be modeled by comparing the observations to
a combination of simulated spectra that have different effective optical
depths. If the line of sight to the quasar passes an ionizing source,
then lines from that region will be shifted to lower column densities
and therefore have little effect on the measured $b(N)$-cutoff (since
it increases with column density). Since the ionizing background
originates from a collection of point sources, minima in the ionizing
background would be shallower and more extended than maxima, except
during reionization. In any case, if the sightline goes through a
region where the mean neutral hydrogen density is substantially
enhanced and therefore affects the measured cutoff, this will become
clear when the spectrum is analyzed in redshift bins.

We conclude that even if local effects
would produce a large scatter in the thermal state of the low-density
gas, the relation between the mean equation of state and the
$b(N)$-cutoff would remain unchanged.

\section{MONTE CARLO SIMULATIONS}

\label{sec:montecarlo}

In section~\ref{sec:algorithm} we described how bootstrap resampling
can be used to reduce the statistical variance in the measured
$b(N)$-cutoff. Given a set of absorption lines from a single QSO
spectrum, the bootstrap method is used to generate synthetic data 
sets, for which the cutoff is measured. The medians of the
resulting probability distributions for the parameters of the cutoff
are then used as best estimates of the true medians.

In order to see how well the equation of state can be measured, we
performed Monte Carlo simulations. We drew 300~random lines, with
column density in the range $10^{12.5}\,\cm^{-2} \le N \le
10^{14.5}\,\cm^{-2}$, from
model~Ob at $z=3$ and used the bootstrap method to generate
probability distributions for the parameters of the cutoff. The median
$\log b_0$ and $\Gamma$ were then converted to measurements of $\log T_0$
and $\gamma$ using the linear relations determined from the
simulations. The statistical variance in the medians of the bootstrap
distributions was estimated from 100 Monte Carlo simulations. For 300
lines per spectrum the dispersion in the median is: $\sigma_{\rm
stat}(\log T_0) = 0.033$~(K), $\sigma_{\rm stat}(\gamma) = 0.13$. 

\begin{figure}
\resizebox{\colwidth}{!}{\includegraphics{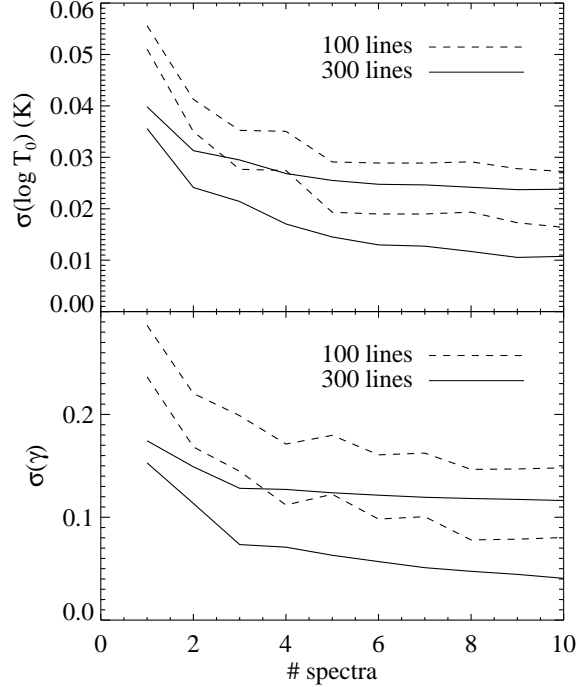}}
\caption{The dispersion in the amplitude (top panel) and slope (bottom
panel) of the equation of state as a function of the number of spectra
used. The dispersion is computed from 100~Monte Carlo
simulations of spectra with 100 (dashed curves) or 300 (solid
curves)~absorption lines each, at $z=3$. The lower curves indicate the
statistical variance, while the upper curves give the
total errors (statistical and systematic). Combining spectra has only
limited effect since systematic errors dominate. The statistical errors
asymptote to the bin size used for determining the probability
distributions.}
\label{fig:montecarlo}
\end{figure}

If multiple spectra are available, the statistical variance in the
median can be reduced by summing the bootstrap probability
distributions of the different spectra. 
Fig.~\ref{fig:montecarlo} illustrates how the errors change
when multiple spectra are used, each containing 100 (dashed curves) or
300 (solid curves) absorption lines in the column density range over
which the cutoff is fitted. The bottom curve of each pair indicates the
statistical $1\sigma$ error. The statistical dispersion asymptotes to
the bin size used for determining the probability distributions.

Besides the statistical variance in the median, there is some
scatter around the linear relations between the parameters of the
cutoff and the equation of state. For example, for 300 lines per
spectrum at $z=3$, the systematic 
errors (i.e.\ the dispersion of the solid circles around the dashed
lines in Figs.~\ref{fig:intercept-T} and \ref{fig:slope-gamma}) in the
parameters of the cutoff are $\sigma_{\rm sys}(\log b_0) = 0.009~(\kms)$ and
$\sigma_{\rm sys}(\Gamma) = 0.015$. The top curve of each pair in
Fig.~\ref{fig:montecarlo} indicates the total (statistical plus
systematic) $1\sigma$ error. 

For a single spectrum with 300~lines (typical for $z=3$),
the predicted total errors are: $\sigma(\log T_0) = 0.040$~(K),
$\sigma(\gamma) = 0.17$. For three spectra
the errors reduce to $\sigma(\log T_0) = 0.029$~(K) and $\sigma(\gamma) =
0.13$. Adding more spectra has very little effect since systematic
errors dominate. 
The errors are larger when there are only 100 absorption lines per
spectrum. In this case the statistical variance is larger and
the errors can be reduced significantly by adding more spectra. The
fact that even for this small number of lines the constraints on the
parameters of the equation of state are significant, suggests that the
method will also work at $z\sim 2$.

\section{SUMMARY AND DISCUSSION}

\label{sec:discussion}

Numerical simulations indicate that the smooth, photoionized
intergalactic medium (IGM) responsible for the low column density
\lya\ forest follows a well defined temperature-density relation. For
densities around the cosmic mean, shock-heating is negligible and the
equation of state of the gas is well-described by a power-law
$T=T_0(\rho/\bar{\rho})^{\gamma-1}$. The equation of state depends on
cosmology, reionization history and the hard X-ray background.
Although the absorption spectra can be fitted by a set of Voigt
profiles, the lines do not in general correspond to discrete
high-density gas clouds. Scatter plots of the distribution of
line widths (\bpar s) as a function of column density ($N$) in
observed QSO spectra clearly show a lower envelope, which increases
with column density. 

The decomposition of spectra produced by a fluctuating IGM into
discrete Voigt profiles is artificial. However, the column density of
the absorption lines correlates strongly with the density of the gas
responsible for the absorption. Although the \bpar s are in general
not correlated with the temperature of the gas, the line widths of the
subset of lines that are close to the $b(N)$-cutoff do show a strong
correlation with temperature. This implies that there exists a lower
limit to the line width, set by the the thermal state of the absorbing
gas, which in turn depends on its density. Hence the cutoff seen in
the $b(N)$-distribution is a direct consequence of the existence of a
temperature-density relation for the low-density gas and can be
used to measure the equation of state of the IGM.

We developed an iterative procedure for fitting a power-law, $b = b_0
(N/N_0)^{\Gamma -1}$, to the $b(N)$-cutoff over a certain column
density range ($10^{12.5}\,\cm^{-2} \le N \le 10^{14.5}\,\cm^{-2}$ at
$z=3$). The algorithm is insensitive to unphysically narrow lines,
which occur in blends and as unidentified metal lines. The intercept
of the power-law, $\log b_0$, can be measured very precisely and is
shown to be very sensitive to $\log T_0$
(Fig.~\ref{fig:intercept-T}). The slope of the cutoff, $\Gamma -1$, is
proportional to $\gamma -1$, but the dependence is weak and it is
harder to measure (Fig.~\ref{fig:slope-gamma}).

The intercept of the $b(N)$-cutoff is a measure of the temperature of
the gas responsible for the absorption lines that are used to
determine the cutoff. This gas is typically slightly overdense and
consequently the intercept depends on both $T_0$, the temperature at
mean density, and $\gamma$, the slope of the equation of
state. However, if we normalise the equation of state to the
temperature at the characteristic density of the gas,
$T=T_\delta(\rho/\rho_\delta)^{\gamma-1}$, where $\rho_\delta \equiv
\bar{\rho}(1+\delta)$, then the intercept of the cutoff depends only
on the amplitude of the equation of state, $T_\delta$.

The relation between the cutoff and the equation of state is
independent of the assumed cosmology (for a fixed equation of
state). In particular, it remains unchanged when all peculiar
velocities are set to zero and when the contribution of the Hubble
flow to the line widths is varied. Changing the effective optical
depth (i.e.\ rescaling the ionizing background), alters the relation
between column density and gas density and thus the relation between
$\log b_0$ and $\log T_0$. However, the dependence is weak and
realistic errors in the measured mean absorption do not lead to
significant errors in the derived value of $T_0$. Variations in the
signal to noise ratio are also unimportant, as long as the ratio is
greater than about 20.

The simulations used to determine the relation between the cutoff and
the equation of state do not incorporate some potentially important
physical processes, like feedback from e.g.\ star formation. This will
however not change the results presented in this paper, as long as the
bulk of the low-density gas follows a power-law equation of
state. Since local effects like feedback would increase the
temperature and since gas cooled to a temperature below that given by
the equation of state is quickly reheated, any additional scatter in
the thermal state of the gas is unlikely to affect the cutoff in the
gas temperature. Doubling the scatter around the equation of state has
no discernible effect on the $b(N)$-cutoff.

The finite number of absorption lines per QSO spectrum introduces
statistical variance in the measured cutoff. The statistical variance
can be reduced by using the bootstrap method to generate probability
distributions for the parameters of the cutoff and using the
medians as the best estimates of the true parameters. If multiple
spectra are available, the variance can be further reduced by adding
the bootstrap distributions of the different spectra.
We use Monte Carlo simulations to estimate the statistical
variance in the medians of the bootstrap distributions. 
Besides the statistical variance, there is a systematic uncertainty from the
scatter in the linear relations between the parameters of the cutoff
and the equation of state. 

For a single spectrum at $z=3$ we predict the following
total (statistical plus systematic) errors: $\sigma(\log
T_0) = 0.040$~(K) ($\Delta T_0 / T_0 = 0.09$), $\sigma(\gamma) =
0.17$.  For three spectra the errors 
reduce to $\sigma(\log T_0) = 0.029$~(K) ($\Delta T_0/T_0 = 0.07$) and
$\sigma(\gamma) = 0.13$. Increasing the number of spectra beyond three
has very little 
effect on the total uncertainty because systematic errors dominate. 
These errors should be compared to the ranges considered to be
physically reasonable, $10^{3.0}\,\K < T_0 < 10^{4.5}\,\K$ and  $1.2 <
\gamma < 1.7$ \cite{hui97:zeldovich_coldensdistr,hui97:tempdens}. 

The analysis presented in this paper is for redshift $z=3$.
At smaller redshifts, the constraints will be less tight because of
the smaller number of lines per spectrum. However, we showed that even
for one third of the number of lines typical at $z=3$, the constraints
on the equation of state are significant. Furthermore, the statistical
variance can be reduced by using multiple spectra.
At higher redshifts the errors will also be somewhat larger than at
$z=3$, mainly because the higher density of lines increases the number
of blends and the errors in the continuum fit.

When using the simulations to convert the observed
$b(N)$-cutoff into an equation of state, one has to be careful to
treat the observed and simulated spectra in the same way. For example,
the simulated and observed $b(N)$-distributions should have the same
number of lines and the same continuum and Voigt profile fitting
algorithms should be used for the simulated and observed spectra. 

Given the existence of many high quality quasar absorption line
spectra, it should be possible to greatly reduce the uncertainty in
the equation of state of the low-density gas over the range
$z=2-4$. This will allow us to put significant constraints on the
reionization history of the universe.

\section*{ACKNOWLEDGMENTS}
We would like to thank M.~Haehnelt and M.~Rauch for
stimulating discussions and R.~Carswell for helping us with \vpfit.
JS thanks the Isaac Newton Trust, St.~John's College and PPARC for
support, AL thanks PPARC for the award of a research studentship and GE
thanks PPARC for the award of a senior fellowship. This work has been
supported by the TMR network on `The Formation and Evolution of Galaxies',
funded by the European Commission.

{}

\end{document}